# What is the effect of EU's fuel-tax cuts on Russia's oil income?[1]

Johan Gars[2]
Daniel Spiro[3]
Henrik Wachtmeister[4]

Date: April 25, 2022

**Abstract**: Following the oil-price surge in the wake of Russia's invasion of Ukraine, many countries in the EU are cutting taxes on petrol and diesel. Using standard theory and empirical estimates, we assess how such tax cuts influence the oil income in Russia. We find that a tax cut of 20 euro cents per liter increase Russia's oil profits by around 11 million Euros per day in the short run and long run. This is equivalent to 4100 million Euros in a year, 0.3% of Russia's GDP or 7% of its military spending. We show that a cash transfer to EU citizens, with an equivalent fiscal burden as the tax cut, reduces these side effects to a fraction.

---

[1] We thank Sarah O'Brien for excellent research assistance and Niklas Rossbach, Per Olsson and Johan Norberg at the Swedish Defence Research Agency for valuable input on Russian military spending and Praman Bansal for important comments.

[2] Beijer Institute, The Royal Swedish Academy of Sciences, Stockholm, Sweden
[3] Dept. of Economics, Uppsala University, Sweden
[4] Dept. of Earth Sciences, Uppsala University, Sweden

# 1 Introduction

Following Russia's military attack on Ukraine, the EU and US have imposed a large number of sanctions on Russia (European Commission 2022, U.S. Dep. of the Treasury 2022). The attack has also led to a negative supply shock of oil, partly since Russia's ability to export has been hampered by the lack of will to insure Russian ships (Reuters 2022a). Consequently, the price of transport fuels has reached record levels (BBC 2022). In response, a large number of European countries are either discussing or have already implemented a reduction in fuel taxes.[5] Such tax reductions have problematic consequences since they increase demand thus making current supply even more scarce. Some of the tax reduction will be attenuated by the underlying oil price increasing, leading to increased profits for oil producers. This paper assesses the magnitude of this effect using basic theory and empirical estimates from the oil sector. We ask: how much does the oil income in Russia increase following fuel-tax reductions in the EU?

# 2 Theoretical approach

We will here describe how we derive the effects of a lowered fuel tax in the EU on Russian oil profits. Our analysis uses a standard model of supply and demand for oil. Our approach is similar to, e.g., Erickson and Lazarus (2014) and Faehn et al (2017). To analyze the effects of the EU's tax, we distinguish between oil demand for road transport fuels in the EU and remaining global oil demand. Similarly, to focus on the effects on Russian revenues, we distinguish between oil supply from Russia and supply from the rest of the world. We will also consider the alternative policy of income transfers to households.

## 2.1 Lowered fuel tax

The first step is to derive how the global oil market responds to changes in the EU road transport fuel tax. The global oil price $p$ is the amount the oil producers receive per unit of oil. To make the oil usable as fuel for end-users, the oil must be refined and transported etc. We assume a globally homogenous per unit cost $c$ for this.[6] Additionally, road fuel users in the EU pay a fuel excise tax $\tau$ per unit of fuel. Furthermore, there are VAT rates $v_{EU}$ and $v_{ROW}$ in the EU and the rest of the world respectively. The market equilibrium is found by equating road fuel demand from the EU, $D_{EU}((1 + v_{EU})(p + c + \tau))$, and remaining global demand for oil products, $D_{ROW}((1 + v_{ROW})(p + c))$, to the supply from Russia $S_{RU}(p)$ and from the rest of the world, $S_{ROW}(p)$:

$$D_{EU}\big((1 + v_{EU})(p + c + \tau)\big) + D_{ROW}\big((1 + v_{ROW})(p + c)\big) = S_{RU}(p) + S_{ROW}(p) \quad (1)$$

---

[5] This includes Austria, Belgium, France, Germany, Italy, Netherlands, Romania, Sweden, UK. See, e.g. Reuters (2022b).
[6] In the appendix we consider the case where these costs are instead proportional to the oil price.

Since our focus here is on tax cuts on transport fuel, $D_{EU}$ should be understood as the demand for oil-based road transport fuel in the EU, i.e., mainly petrol and diesel. With some abuse of technical differences, we will often refer to it only as fuel. $D_{ROW}$ should be understood as the global demand for all other oil products except road fuel in the EU, that is, it includes also non-road oil products in the EU.

The effect of a change in the tax on the equilibrium price can then be found by treating the price as a function of the tax, differentiating the equilibrium condition fully with respect to the tax, solving for the derivative of the price with respect to the tax and rewriting. Let $x$ denote the share of global demand for oil products that comes from road fuel demand in the EU and let $y$ denote Russia's share of the global oil supply. The market response to a change in the tax depends on the supply and demand elasticities in the submarkets. Let $\varepsilon_{D,EU}$, $\varepsilon_{D,ROW}$, $\varepsilon_{S,RU}$ and $\varepsilon_{S,ROW}$ denote the demand elasticities in the EU and the rest of the world and the supply elasticities of Russia and the rest of the world respectively.[7] Using this notation, the effect of a tax change on the oil price is given by the derivative[8]

$$\frac{dp}{d\tau} = \frac{x \frac{p}{p+c+\tau} \varepsilon_{D,EU}}{y \varepsilon_{S,RU} + (1-y)\varepsilon_{S,ROW} - x \frac{p}{p+c+\tau} \varepsilon_{D,EU} - (1-x)\frac{p}{p+c}\varepsilon_{D,ROW}}. \tag{2}$$

Note the ratios multiplying the demand elasticities. These ratios correct for the fact that demand depends on the price including the supply costs and taxes and hence that a change in the global oil price $p$ will have a smaller percentage effect on consumer prices. We can also see that the VAT rates cancel from these expressions. The VAT rate in the EU shows up in other expressions below, but the results are independent of the VAT rate in the rest of the world.

The effect of a tax change $\Delta \tau$ on the oil price can be linearly approximated as

$$\Delta_\tau p \approx \frac{dp}{d\tau} \Delta \tau. \tag{3}$$

Let the fuel price in the EU be denoted $f \equiv (1 + v_{EU})(p + c + \tau)$. The change in the fuel price is

$$\Delta_\tau f \approx (1 + v_{EU})(\Delta_\tau p + \Delta \tau). \tag{4}$$

The EU's fuel tax revenues are $T_{EU} \equiv (v_{EU}(p+c) + (1+v_{EU})\tau)D_{EU}$, and the fiscal burden of the tax change can be found by differentiating with respect to $\tau$, taking into account that the oil price depends on the tax, and making a linear approximation:

$$\Delta_\tau T_{EU} \approx \left[1 + \left(v_{EU} + \frac{\tau + v_{EU}(p+c+\tau)}{p+c+\tau}\varepsilon_{D,EU}\right)\left(1 + \frac{dp}{d\tau}\right)\right] D_{EU} \Delta \tau. \tag{5}$$

Finally, we will translate an oil price change into a change in Russian oil profits. The oil profits are $\pi_{RU} = (p - e)S_{RU}(p)$, where $e$ represents oil extraction costs which are assumed to be constant.

---

[7] The demand elasticities measure responses in demanded quantities to changes in the end-user prices (including additional costs and taxes) while supply elasticities measure responses of supplied quantities to changes in the oil price.
[8] When rewriting the derivative, we use that $D_{EU} + D_{ROW} = D = S = S_{RU} + S_{ROW}$, and the relations $D_{EU} = xD$, $D_{ROW} = (1-x)D$, $S_{RU} = yS$ and $S_{ROW} = (1-y)S$.

This assumption is realistic under the production changes considered here. Again, treating $p$ as a function of $\tau$, differentiating with respect to $\tau$, and making a linear approximation gives that

$$\Delta_\tau \pi_{RU} \approx \left(1 + \frac{p-e}{p}\varepsilon_{S,RU}\right) S_{RU}(p)\Delta_\tau p. \tag{6}$$

## 2.2 Income transfers

An alternative way to compensate people for the increase in the fuel cost is to give general income transfers to people corresponding to the reduction in tax revenues that would result from a decreased fuel tax. From a welfare perspective, this is preferable since people can then choose how to use the money. From the perspective of this paper, an additional benefit is that a share smaller than one will be spent on fuel and hence that the increase in Russian oil profits will be smaller. How much smaller will be assessed quantitatively.

To analyze this alternative policy option, we assume that road fuel demand in the EU now depends on disposable income $I$ in addition to the fuel price. Differentiate the market equilibrium condition with respect to $I$ and treat the equilibrium price as a function of income to get

$$\frac{dp}{dI} = \frac{p}{I} \frac{x\varepsilon_{I,EU}}{y\varepsilon_{S,RU} + (1-y)\varepsilon_{S,ROW} - x\frac{p}{p+c+\tau}\varepsilon_{D,EU} - (1-x)\frac{p}{p+c}\varepsilon_{D,ROW}}, \tag{7}$$

where $\varepsilon_{I,EU}$ is the income elasticity of road fuel demand in the EU. The effects of a change in the disposable income $\Delta I$ on the oil price, $p$, and the fuel price in the EU, $f$, are

$$\Delta_I p \approx \frac{dp}{dI}\Delta I \text{ and } \Delta_I f \approx (1 + v_{EU})\Delta_I p. \tag{8}$$

The effect on Russian oil profits due to the income transfer is

$$\Delta_I \pi_{RU} \approx \left(1 + \frac{p-e}{p}\varepsilon_{S,RU}\right) S_{RU}\Delta_I p. \tag{9}$$

## 3 Three cases

We quantitatively assess the effects described theoretically in three different cases: the very short run, short run and long run. We here sketch the qualitative effects in the different time horizons. The size of the effects obviously depends on the numerical values (see next section). We start by describing the long run effects.

In the long run (LR), to be thought of as beyond one year, supply is somewhat elastic, and demand is rather elastic too. This is since producers have time to adjust their production and make some capacity investments. Likewise, consumers can acquire new habits or find solutions based on a new fuel price and those in the process of buying a new vehicle will be affected by the fuel price (see e.g., Severen and van Benthem 2022). This is the case roughly sketched in Figure 1. The grey curves are demand from the EU and the ROW respectively. The black curves are global demand and supply. A tax reduction in the EU shifts the EU's demand outwards (the dashed grey curve), this in turn increases global demand with an equivalent amount (the dashed black curve). There are two effects of this: an increase in the oil price (a shift along the vertical axis) and an increase in the quantity of oil produced (a shift along the horizontal axis).

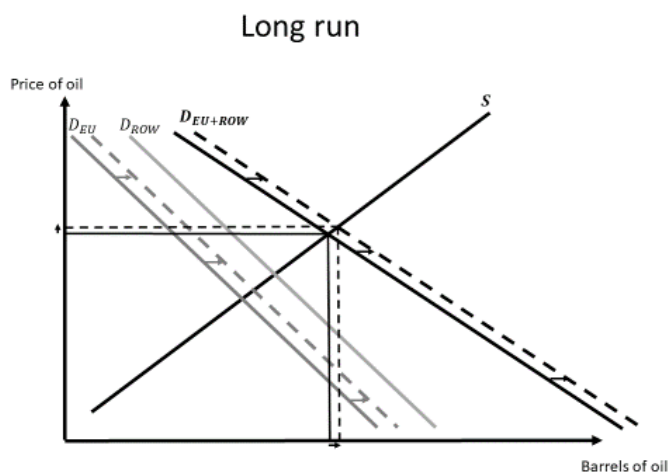

*Figure 1. Illustration of supply and demand in the long run.*

In the short run (SR), to be thought of as 1-12 months, the supply elasticity in terms of quantity is lower than in the long run. This can be seen in Figure 2 (a) where supply is illustrated as fixed. Demand elasticity is lower than in the long run since most of the consumer choice regards how much to drive rather than what vehicle to buy or how to change long-term habits. Since, as in the figure, supply is fixed, a reduced tax results in increased oil price but no increase in production or consumption. In practice some of the increased demand from the EU is attenuated by decreased consumption in ROW.

In the very short run (vSR), to be thought of as up to a month, there are limitations on how much oil, originally meant for other markets, that can be quickly redirected to the EU. The reason for this is that bilateral contracts of supply can be viewed as partly fixed – oil tankers on their way to one country cannot in the very short run easily be sent elsewhere. To capture this, we model this *as if* the EU is an isolated oil market. The supply is fixed, both in terms of quantity and in who supplies the oil. Implicitly this also means that the oil price in the EU may differ from the oil price elsewhere. The case is illustrated in Figure 2 (b). Importantly, in this case, Russia is a much larger supplier than on the global market. Demand is also less elastic than in the short run. We view our modeling here more as a thought experiment, reality in the very short run probably lies somewhere in between our modeling here and the short-run illustration above. One simplification and limitation of our model is that we do not consider oil inventories. It should also be noted that if the EU imposes an import ban on Russian oil, the tax effects on Russian income in the very short run disappear.

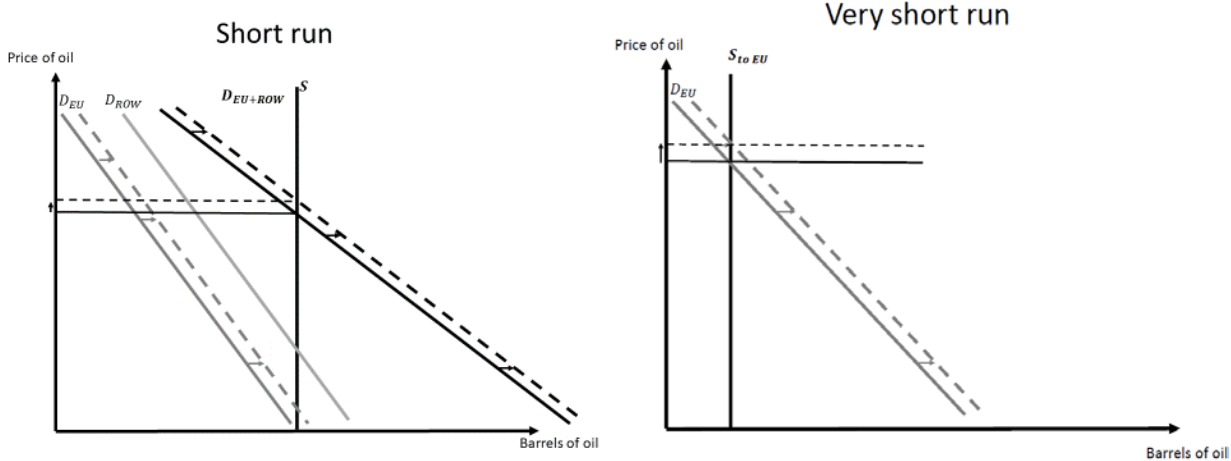

*Figure 2. Illustration of supply and demand in (a) the short run and (b) very short run.*

# 4 Data and estimates

The parameters, quantities and shares in equations 1-9 are based on previous research and current data. They are reported in Table 1. Where relevant, we distinguish between very short-run, short-run and long-run elasticities and shares. Motivations for the values are provided in the table with further information available in the appendix.

In a sensitivity analysis we perturb the key parameters to show how that affects our results, see Appendix.

*Table 1. Parameters for equations 1-9 describing Russia's oil income.*

| Parameter | | V. Short run | Short run | Long run | Reference |
|---|---|---|---|---|---|
| Demand elasticity of road fuel in EU with respect to price | $\varepsilon_{D,EU}$ | -0.25 | -0.25 | -0.9 | See Appendix |
| Demand elasticity of oil globally (excl EU road fuel) with respect to price | $\varepsilon_{D,ROW}$ | -0.125 | -0.125 | -0.45 | See Appendix |
| Supply elasticity of Russian oil with respect to price | $\varepsilon_{S,RU}$ | 0 | 0 | 0.13 | See Appendix |
| Supply elasticity global oil (excl Russia) with respect to price | $\varepsilon_{S,ROW}$ | 0 | 0 | 0.13 | See Appendix |
| Demand elasticity of road fuel in EU with respect to income | $\varepsilon_{I,EU}$ | 1 | 1 | 1 | Dahl (2012) |
| Base fuel excise tax EU | $T$ | 0.6 EUR/l | 0.6 EUR/l | 0.6 EUR/l | Tax Foundation (2021) |
| Base oil price per liter fuel | $p$ | 0.58 EUR/l | 0.58 EUR/l | 0.58 EUR/l | Oil price $110/b; FX-rate 0.9 €/$; 170 l fuel/b |
| Other costs (refining, etc) | $c$ | 0.48 EUR/l | 0.48 EUR/l | 0.48 EUR/l | Follows from $p + \tau + c$, see Appendix |
| VAT rate | $v_{EU}$ | 0.2 | 0.2 | 0.2 | |
| Base consumer price | $p + \tau + c$ | 2.0 EUR/l | 2.0 EUR/l | 2.0 EUR/l | Sum of oil price, fuel tax and other costs |
| Oil extraction cost Russia per barrel | $e$ | 17 EUR/b | 17 EUR/b | 17 EUR/b | Rystad UCube database |
| EU road fuel demand share of global oil consumption | $x$ | 48% | 5.7% | 5.7% | BP (2021), Eurostat (2022) |
| Russian oil exports (liters fuel and barrels of oil) | $S_{RU}$ | 440 Ml/d  2.6 Mb/d | 940 Ml/d  5.5 Mb/d | 1 400 Ml/d  8 Mb/d | IEA (2022) and case assumptions, see Appendix |
| Global oil supply, excl Russian exports (liters fuel and barrels of oil) | $S_{RoW}$ | 13 000 Ml/d  7.6 Mb/d | 16 000 Ml/d  92 Mb/d | 16 000 Ml/d  92 Mb/d | Follows from $S_{RU}, S$ |
| Road fuel demand in EU (liters fuel and barrels of oil) | $D_{EU}$ | 820 Ml/d  4.8 Mb/d | 950 Ml/d  5.6 Mb/d | 970 Ml/d  5.7 Mb/d | Follows from $x, D$ |
| Global oil demand, excl EU road fuel demand (liters fuel and barrels of oil) | $D_{ROW}$ | 910 Ml/d  5.4 Mb/d | 16 000 Ml/d  92 Mb/d | 16 000 Ml/d  94.3 Mb/d | Follows from $x, D$ |
| Russia's exports as share of global supply | $y$ | 25% | 5.60% | 8% | Follows from $S_{RU}, S$ |

# 5   Results and discussion

## 5.1   Tax cut

We analyze the effects of a fuel-tax cut in the very short run, short run and long run. The tax cut considered is equivalent to 20 euro cents incl. VAT.[9]

The results in the **very short run** are presented in the upper row of Table 2. 9 cents of the tax reduction of 20 cents is passed through to oil suppliers. Russia, being a significant supplier, attains a large share of the fiscal cost of the policy, making an additional 39 million EUR per day. Apart from financing Russia, the policy is also quite ineffective in lowering consumer prices in the very short run – consumers only experience 9 cents of reduction per liter despite the tax reduction being 20 cents. The results here take into account Russia's current reduced supply to the EU (see Table 4 in the appendix). Should the EU impose an import ban on Russian oil, all effects on Russia's income in the very short run disappear.

In the **short run**, the consumer price in the EU is reduced by almost the full tax reduction and the, now global, oil price is increased by much less than in the very short run. Nevertheless Russia, a large supplier also globally, is still getting sizable additional profits – 11 million EUR per day or 4.1 billion EUR in year equivalents.

In the **long run**, to be thought of as beyond one year, supply becomes somewhat elastic, and so does demand. The price effects are again smaller, and the fiscal cost to the EU is smaller than in the very short and short run. This is because, instead of an oil-price increase, there is an increase in the supply. Russia's additional oil profits are still sizeable at 12 million EUR per day or 4.3 billion EUR per year. The main reason that Russia's profits are larger in the long run than in the short run is because we assume that the currently reduced exports (5.5 Mb/d) will rebound back to normal (8 Mb/d) in the long run.

*Table 2. Effect of an EU fuel tax cut of 20 euro cents/liter.*

|  | Oil price change cents/liter | EU fuel price change cents/liter | Fiscal cost EU MEUR/day | Profit gain Russia MEUR/day | Profit gain Russia MEUR/year |
|---|---|---|---|---|---|
| Very short run | 8.9 | -9.3 | 140 | 39 | – |
| Short run | 1.2 | -18 | 170 | 11 | 4 100 |
| Long run | 0.79 | -19 | 115 | 12 | 4 300 |

In the appendix we carry out a sensitivity analysis of these results with respect to other parameter values. The results are generally not very sensitive in the long and very short run. The short-run results are more sensitive. The sensitivity analysis suggests Russia's short-run profit gain can be a third compared to those using our preferred parameter values (reported here in the main text). But they may also be twice as high. We also investigate whether the discount on Russian oil (through the Urals price) is likely to change our results and whether letting the costs $c$ be proportional to the oil price.

---
[9] That is, the cut in the fuel duty tax is 0.2/(1+VAT) EUR.

Are these additional profits for Russia large? We now discuss this from a few different perspectives. First note that in the vSR, a large share of the EU's fiscal cost, 28%, is sent to Russia. In the SR and LR much less so but still around 5-10% of what is meant to help European consumers is instead helping Russia.

The additional Russian profits are sizeable compared to Russia's pre-invasion GDP which was about 3.7 billion EUR per day. The EU's tax cut increases Russia's GDP by around 1% in the vSR, 0.3% in the SR and the LR. We can also compare them to Russia's military spending, which pre-invasion was about 160 million EUR per day.[10] The daily profit increase then corresponds to 24%, 7% and 8% of military spending in the vSR, SR and LR respectively.

Russia's government income is to a large extent accrued from oil (fluctuating between 30 and 50% in normal times, see e.g. Sabitova and Shavaleyeva 2015) and it now, under credit constraints, has massive outlays on military and repression equipment and services. It is hard to say how much of an increased income will be invested in military capacity and repression but one can still get a sense of how much the additional income *could* buy in such capacities. Based on various informal sources (see Table 5 in the appendix), in Russia, contract soldiers receive the equivalent of 7500 EUR per year, entry-level policemen earn around 7200 EUR per year and a troll-farm worker receives 6800 EUR per year. One day of the short-run income of 11 million EUR thus allows for covering the *yearly* salary of over 1400 contract soldiers fighting in Ukraine, or 1500 policemen arresting demonstrators and controlling the media, or 1600 disinformation workers.

It is difficult to know the price of military equipment in Russia. But the US-made M270 MLRS, a modern equivalent of the BM-21 Grad (which Russia is using for shelling civilian quarters in Ukraine) costs around 2 million EUR. One day of additional short-run income is enough to equip Russia with 5 such vehicles. Upgrading the T-72B main battle tank to modern standard may cost as little as 211000 EUR. One day of additional short-run income is thus enough to modernize over 50 tanks.[11]

## 5.2 Fiscal equivalent cash transfer

As seen, a fuel-tax cut in the EU has substantial downsides. It provides Russia with a large additional income. Furthermore, part of the help meant for EU fuel consumers in the form of a tax cut is instead passed through to increased oil prices, especially in the very short run. The question is then whether it is possible to help EU consumers in a way that does not benefit Russia. We here look at one such alternative, namely in the form of a cash transfer to consumers with an equivalent fiscal burden as the tax cut. That is, we tie the transfer to 140 MEUR, 170 MEUR and 115 MEUR in the very short, short and long run respectively.

---

[10] Based on World Bank (2022), the average yearly military spending 2015-2020 was 65 billion USD. A USD/EUR exchange rate of 0.9 makes this 160 million EUR per day.
[11] Based on various informal sources (see Table 5 in the appendix).

The results are presented in Table 3. The increased income leads to an increased demand for fuels. However, since the cash can be spent on anything, most of it is used for other things than fuels. Hence, the price increases only marginally (1.4 cents in vSR, much less in the long run). One perspective on this, which highlights the main benefit of cash transfers compared to fuel tax cuts, is that consumers, through these policies, get the equivalent of 20 cents for each liter of fuel consumed on average. To get the benefits of a tax cut, a consumer has to buy fuel. If they, instead, get it in the form of cash, they can choose to spend it all on fuel, spend none of it on fuel or somewhere in between – cash is a more flexible currency than a tax cut. Even a person that spends the whole transfer on fuel will gain from a cash transfer since the oil and fuel price increase only by 1.4 cents. This thus allows for varying preferences in the population.[12]

Perhaps most importantly for the subject matter here, Russia's profit gains are substantially lower with the cash transfer, in the short run around a sixth of the profits received from a tax cut. In the long run much less.

*Table 3. Effect of a fiscal equivalent cash transfer.*

|  | Oil price change cents/liter | EU fuel price change cents/liter | Fiscal cost EU MEUR/day | Profit gain Russia MEUR/day | Profit gain Russia MEUR/year |
| --- | --- | --- | --- | --- | --- |
| Very short run | 1.2 | 1.4 | 140 | 5.3 | – |
| Short run | 0.19 | 0.23 | 170 | 1.8 | 650 |
| Long run | 0.024 | 0.029 | 115 | 0.36 | 132 |

---

[12] It is also possible to direct the cash to particular groups that are hit harder by the price increase.

# 6   Conclusion

We have analyzed how much a fuel-tax cut in the EU will increase Russian income from oil. The effects are substantial. A tax cut of 20 euro cents increases Russia's daily oil income with 11 million EUR during a first year and 12 million EUR per day if it remains longer (in the very short run, the daily income increase can be substantially higher). Given Russia's current need for military capacity and lack of resources due to sanctions, the additional income is large enough to enable a substantial increase in Russia's military and repressive capabilities.

We show that a fiscally equivalent cash transfer can achieve similar alleviation to consumers as a tax cut but with a fraction of the increased income to Russia.

# 7 Appendix

The appendix consists of two parts: a more detailed motivation for our parameter values and a sensitivity analysis of our results.

## 7.1 Parameter values

### 7.1.1 Elasticity of demand

The price elasticity of demand for road transport fuels (gasoline and diesel) and crude oil is low in the short term but increases with time. In the short term, many fuel consumers can only drive less to reduce consumption while in the longer term many can shift to more efficient vehicles, change commuting distance or mode of transportation.

Several studies compile existing estimates of demand elasticity of gasoline, diesel and crude oil (see for example Hamilton 2009, Caldara et al 2016, Hössinger et al 2017, Aklilu 2020). These estimates are derived using different methods over different time periods and locations. Short term gasoline elasticity estimates range from -0.04 to -0.5 (Hössinger et al 2017, Aklilu 2020) with several review studies deriving averages around -0.25 (Aklilu 2020). Aklilu (2020) also provides additional original estimates for EU countries using recent data, finding an EU average short-term gasoline elasticity of -0.255. We use -0.25 in our calculations, in line with these recent EU estimates as well as the wider literature samples.

Long term gasoline demand elasticity estimates range from -0.2 to -1 (Hössinger et al 2017). Aklilu (2020) compiles review studies with even higher ranges but with averages around -0.7. Aklilu's (2020) own empirical study finds a long-term EU average of -0.88. We use -0.9 in our calculations based on these results.

Crude oil demand elasticity is usually found to be lower than gasoline. This is to be expected since crude oil is only a part of the gasoline price. If we assume crude oil represents half the cost of retail gasoline, a 10% increase in the price of crude would translate into a 5% increase in the price of gasoline, and the demand elasticities for oil would be about half those for gasoline (Hamilton 2009). Caldara et al (2016) compile 31 studies for short term world oil demand in the range of -0.04 to -0.9 with a mean of -0.22 and a median of -0.13. We follow Hamilton (2009) and use half of the gasoline estimates for wider oil demand, i.e. -0.125 for our short-term oil demand elasticity and -0.45 for our long term, which also is in line with the estimates of Caldara et al (2016).

In the sensitivity analysis we also explore lower and higher values in the range found in the literature.

### 7.1.2 Elasticity of supply

The price elasticity of global oil supply is low, close to zero, in the short term and grows only slowly in the longer term. New conventional oil fields take several years to bring into production and additional supply in the short term (within one year) can only come from either inventory, "politically" withheld supply, shale oil production and infill drilling in conventional fields.

Compared to demand elasticity studies, supply elasticity studies are rare. Caldara et al (2016) compiles 6 studies applying different methods and data and find short term (within one year) supply elasticities in the range of 0-0.27.

For our very Short run and Short run scenario we use a supply elasticity of 0 for both Russia and the rest of the world. This corresponds to a hypothetical scenario with no inventory draw, no additional OPEC production and a timeframe below 6 months where the shale response is still low. In the sensitivity analysis we present a short-run case using 0.05 which can be seen as reflecting either a 12-month shale response or a quicker OPEC response.

For our Long run scenario, we use 0.13 for the rest of the World, which is used by Erickson and Lazarus (2014). A supply elasticity of 0.13 is also in line with a modeled 3-year horizon estimate of global supply by Wachtmeister (2020). In the sensitivity analysis we explore 0.2 as a higher estimate, reflecting either a stronger supply response or a response a few years further ahead.

### 7.1.3 Other costs and refinery margins

We translate oil production (crude oil, condensate, natural gas liquids) in barrels per day to a corresponding volume of refined products. We make the simplifying assumption that 1 barrel of oil yields 170 liters of products and fuels that can be sold to consumers. In the base case, the variable production cost of these fuels is assumed to correspond directly to the crude oil price, i.e. variable fuel production cost is the global oil price per barrel (Brent, in USD/bbl) measured in Euro per liter fuel product ($p$). The retail fuel price (the consumer price) is then variable fuel production cost (oil price $p$) plus fixed other production costs $c$ (refining, transport, margins etc.) plus fuel tax $T$, then VAT is applied to all these. In section 7.2.2 in the sensitivity analyses we explore a variable other production cost ($z$) and discuss which case is more likely. Our base case $c$ of 0.48 EUR/liter is derived "backwards" from a consumer price of 2 EUR/liter, $c$ thus includes the current refinery margins (the value difference between crude oil and refined products) which varies in time and currently are at historically high levels.

### 7.1.4 Size of markets and Russian export declines

We assume Russia has already lost 2.5 Mb/d of oil exports based on recent news and industry reporting (see Table 4). In January, before the war, the global oil supply and demand of oil was estimated to be 100 Mb/d (IEA 2022b). Consequently, we assume the current oil market to be 97.5

Mb/d. Road-transport fuel's share of total EU oil consumption is 47.5% (Eurostat 2022). The EU's share of global oil consumption is 12% (BP 2021) which yields our *x=0.475\*0.12=5.7%*. The EU imports, in normal times, about 35% of its oil from Russia. For the very short run we assume that the reduction in Russia's export (2.5 Mb/d) mainly accrues to EU, i.e., 1.5 Mb/d less than normal. This implies *y=25%* in the very short run.

*Table 4. Estimates of current Russian export declines.*

| Source | Date | Total exports | Crude oil | Oil products |
|---|---|---|---|---|
| Bloomberg (2022) | March 28 | Not disclosed | -1.4 Mb/d | Not disclosed |
| Rystad Energy (2022) | March 22 | Not disclosed | -1.5 Mb/d | Not disclosed |
| IEA (2022c) | March 16 | -2.5 Mb/d | -1.5 Mb/d | -1 Mb/d |
| Energy Intelligence (2022) | March 7 | -3 Mb/d | -1.6 Mb/d | -1.4 Mb/d |

### 7.1.5 Salaries and costs in Russia

We use here an exchange rate of 1 EUR to 80 RUB. Official and reliable statistics of salaries for the mentioned labor categories are scare, besides Oxenstierna (2016) we also rely on some informal sources presented in Table 5. All web-sources accessed April 6, 2022.

*Table 5. Estimates of Russian salaries and material.*

| Source | Category | Cost |
|---|---|---|
| Oxenstierna (2016) | Soldier | 30 000 RUB/month |
| Russian Defense Policy (2021) | Soldier | 30 000 RUB/month |
| EU Today (2022) | Soldier | 62 000 RUB/month |
| Tylaz (2022) | Soldier | 30 000 RUB/month |
| ERI (2022) | Police, entry level | 660 000 RUB/year |
| Time (2015) | Disinformation worker | 45 000 RUB/month |
| ArmedForces (2022) | Rocket artillery, modern version of BM-21 Grad | 2 000 000 EUR/piece |
| Majumdar (2016) | Upgrade of T-72B main battle tank | 211 000 EUR/piece |

## 7.2 Sensitivity analysis

The sensitivity analysis consists of three parts. The first part perturbs the parameter values, including supply from Russia. The second part lets the costs be proportional to the oil price. The third part investigates the potential decoupling between Russia's oil price (Urals) and the global oil price (Brent).

### 7.2.1 Parameter perturbation

With many studies finding estimates of short-run demand elasticity of gasoline around -0.25 we use a relatively narrow spread of ±20% around our base case for $\varepsilon_{D,EU}$, instead of for example using the full range in the literature. The same reasoning is applied for LR. For vSR, we use a slightly lower $\varepsilon_{D,EU}$ minimum value (-0.1).

Oil demand elasticity for the rest of the world is set to be EU's fuel demand elasticity divided by either 1.5 or 3 (instead of 2 as in the baseline). This is in line with the empirical evidence which usually finds oil demand elasticity to be lower than fuel elasticity, as well as being in line with theoretical expectations since crude oil is a share lower than one of the fuel price.

For short-run supply elasticity, we allow a maximum value of 0.1, which would yield a supply response of 2 Mb/d globally at a 20% price increase, and a 5 Mb/d response at a 50% price increase. 5 Mb/d must be deemed as a practical limit of what the global oil supply system can deliver within one year. It could only materialize if all parts of supply were pulling in the same direction with maximum OPEC and Non-OPEC production increases, including maximum shale expansion, OPEC spare capacity, Iran and Venezuela returning from sanctions and large inventory draws. This would not happen by market forces alone triggered by a 50% oil price increase. The 0.1 case should therefore be treated as an extreme political-economic scenario.

In the very short-run case, the distribution of the -2.5 Mb/d of Russian export falls must be specified. In the base case we assume -1.5 Mb/d of exports to Europe and -1 Mb/d of exports to the rest of the world. In the low case, we assume that the full 2.5 Mb/d reduction is applied to exports to Europe. In the Short-run case we also use 8 Mb/d as a high value, accounting for the uncertainty of the actual current export falls. In the Long run we keep Russian export fixed at 8 Mb/d, representing the ability to re-route western export flows to other importers.

*Table 6. Parameter values in sensitivity analysis.*

|  | $\varepsilon_{D,EU}$ | $\varepsilon_{D,ROW}$ | $\varepsilon_{S,RU}$ | $\varepsilon_{S,ROW}$ | $S_{RU}$ |
|---|---|---|---|---|---|
| Very short run | | | | | |
| *Low* | -0.1 | $\varepsilon_{D,EU}/3$ | 0 | 0 | 1.595 Mb/d |
| *Base case* | -0.25 | -0.125 | 0 | 0 | 2.595 Mb/d |
| *High* | -0.3 | $\varepsilon_{D,EU}/1.5$ | 0 | 0 | 4.095 Mb/d |
| Short run | | | | | |
| *Low* | -0.2 | $\varepsilon_{D,EU}/3$ | 0 | 0 | 5 Mb/d |
| *Base case* | -0.25 | -0.125 | 0 | 0 | 5.5 Mb/d |
| *High* | -0.3 | $\varepsilon_{D,EU}/1.5$ | 0.1 | 0.1 | 8 Mb/d |
| Long run | | | | | |
| *Low* | -0.7 | $\varepsilon_{D,EU}/3$ | 0.05 | 0.05 | 8 Mb/d |
| *Base case* | -0.9 | -0.45 | 0.13 | 0.13 | 8 Mb/d |
| *High* | -1.1 | $\varepsilon_{D,EU}/1.5$ | 0.2 | 0.2 | 8 Mb/d |

The upper- and lower-bound values are presented in Table 6. To test sensitivity we look at the outcomes combining all upper and lower bounds. As there are five parameters, each with two values, this totals 32 outcomes for each of the time horizons.

*Table 7. Results of sensitivity analysis.*

|  | Oil price change euro cents/liter | EU fuel price change euro cents /liter | Fiscal cost EU MEUR/day | Profit gain Russia MEUR/day |
|---|---|---|---|---|
| Very short run | | | | |
| *Minimum* | 7.7 | -11 | 170 | 21 |
| *Base case* | 8.9 | -9.3 | 140 | 39 |
| *Maximum* | 11 | -7.3 | 130 | 74 |
| Short run | | | | |
| *Minimum* | 3.8 | -20 | 180 | 3.4 |
| *Base case* | 1.2 | -19 | 170 | 11 |
| *Maximum* | 1.7 | -18 | 160 | 24 |
| Long run | | | | |
| *Minimum* | 0. 51 | -19 | 130 | 7.4 |
| *Base case* | 0. 79 | -19 | 115 | 12 |
| *Maximum* | 1.4 | -18 | -97 | 21 |

The results of the multiple combinations are presented numerically in Table 7 and graphically in Figures 3-5 for the vSR, SR and LR. The table presents the highest and lowest result attained among all parameter combinations. Generally, the SR results are more sensitive than vSR and LR. The smallest increase in Russian profits is a third of our baseline case and the largest is about twice as large. In the vSR and LR it is particularly the downside variation that is smaller. The fiscal cost to the EU does not vary much for any combination. The oil price change is somewhat sensitive under the most extreme combinations, mainly downwards in SR and mainly upwards in LR.

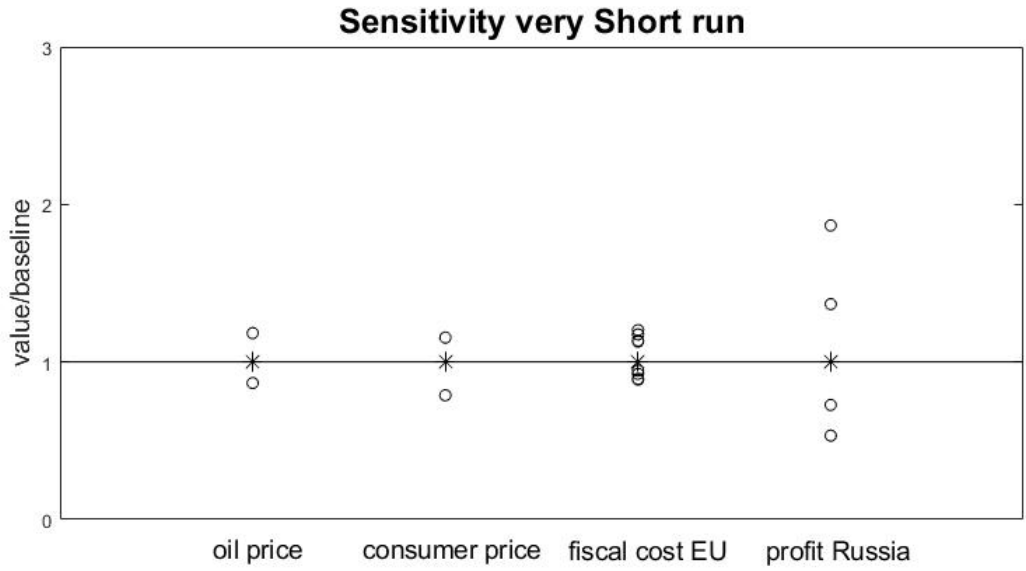

*Figure 3. Sensitivity analysis very short run.*

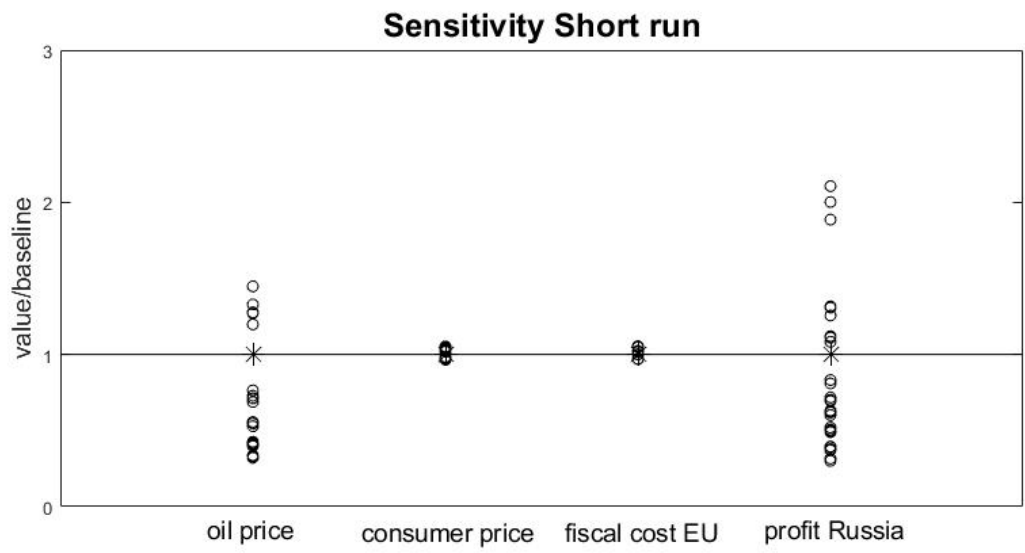

*Figure 4. Sensitivity analysis short run.*

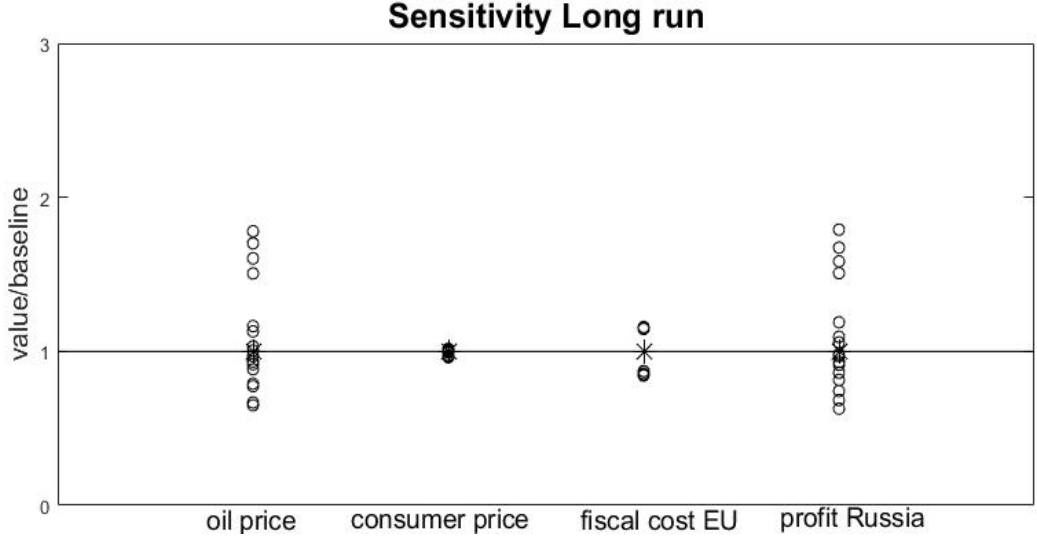

*Figure 5. Sensitivity analysis long run.*

### 7.2.2 Proportional instead of per-unit refinement and transportation costs

In the baseline case we have assumed that the cost $c$ for refinement, transportation etc. is a per-unit cost and independent of the oil price. It is, however, plausible that part of this cost would increase when the oil price goes up, for instance, since refining and transportation of oil and fuel in itself uses fuel. This would make fuel demand more sensitive to the oil price and we expect the effects on Russian oil profits to be smaller. The most extreme form of that would be that the entire cost is proportional to the oil price. We here investigate that case and denote the proportional cost by $z$. The fuel prices would then be $(1+v_{EU})((1+z)p+\tau)$ and $(1+v_{ROW})(1+z)p$, instead of $(1+v_{EU})(p+c+\tau)$ and $(1+v_{ROW})(p+c)$ in the EU and the rest of the world respectively.

Following similar derivation as for the baseline model, the effect of a change in the tax on the equilibrium price becomes

$$\frac{dp}{d\tau} = \frac{x\frac{p}{(1+z)p+\tau}\varepsilon_{D,EU}}{y\varepsilon_{S,RU}+(1-y)\varepsilon_{S,ROW}-x\frac{(1+z)p}{(1+z)p+\tau}\varepsilon_{D,EU}-(1-x)\varepsilon_{D,ROW}}$$

while the linear approximation of the change in the price is still

$$\Delta_\tau p \approx \frac{dp}{d\tau}\Delta\tau.$$

The fuel price in the EU is now $f=(1+v_{EU})((1+z)p+\tau)$. And the change in it is

$$\Delta_\tau f \approx (1+v_{EU})((1+z)\Delta_\tau p+\Delta\tau).$$

The EU's fuel tax revenues are $T_{EU} \equiv (v_{EU}(1+z)p + (1+v_{EU})\tau)D_{EU}$, and the fiscal burden of the tax change becomes

$$\Delta_\tau T_{EU} \approx \left[1 + \left(v_{EU} + \frac{\tau + v_{EU}((1+z)p + \tau)}{(1+z)p + \tau}\varepsilon_{D,EU}\right)\left(1 + (1+z)\frac{dp}{d\tau}\right)\right]D_{EU}\Delta\tau.$$

The expression for the change in Russian oil profits is the same as before.

Considering the effect of a change in the disposable income in the EU, the effect on the oil price now becomes

$$\frac{dp}{dI} = \frac{p}{I}\frac{x\varepsilon_{I,EU}}{y\varepsilon_{S,RU} + (1-y)\varepsilon_{S,ROW} - x\frac{(1+z)p}{(1+z)p+\tau}\varepsilon_{D,EU} - (1-x)\varepsilon_{D,ROW}}.$$

The effects of a change in the disposable income $\Delta I$ on the oil price, $p$, and the fuel price in the EU, $f$, are now

$$\Delta_I p \approx \frac{dp}{dI}\Delta I \text{ and } \Delta_I f \approx (1+v_{EU})(1+z)\Delta_I p.$$

The effect on Russian oil profits due to the income transfer is the same.

For our numerical computations, we set the proportional cost so that the resulting EU fuel price is equal to 2 EUR, the same as in the baseline. This gives a value $z = 0.83$.

*Table 8. Effect of an EU fuel tax cut of 20 euro cents/liter with proportional costs.*

|  | Oil price change cents/liter | EU fuel price change cents/liter | Fiscal cost EU MEUR/day | Profit gain Russia MEUR/day | Profit gain Russia MEUR/year |
|---|---|---|---|---|---|
| Very short run | 4.9 | -9.3 | 141 | 22 | – |
| Short run | 0.65 | -19 | 166 | 6.1 | 2 200 |
| Long run | 0.51 | -19 | 115 | 7.6 | 2 800 |

*Table 9. Effect of a fiscal equivalent cash transfer with proportional costs.*

|  | Oil price change cents/liter | EU fuel price change cents/liter | Fiscal cost EU MEUR/day | Profit gain Russia MEUR/day | Profit gain Russia MEUR/year |
|---|---|---|---|---|---|
| Very short run | 0.66 | 1.4 | 141 | 2.9 | – |
| Short run | 0.10 | 0.23 | 166 | 0.97 | 350 |
| Long run | 0.016 | 0.034 | 115 | 0.23 | 85 |

The effects of the tax reduction are given in Table 8 and the effects of the income increase are given in Table 9. Comparing these to the baseline results, we can see that the effects on the oil price and Russian profits are roughly half as large with proportional compared to additive costs. This is reasonable since the proportional cost means that any change in the oil price will be amplified by a factor of about 1.8 with the proportional costs.

What is plausible to assume about the costs, are they proportional to the oil price or fixed? It is hard to say with precision. Most likely it is somewhere in between, so the two extreme cases of constant costs (baseline) and fully proportional costs (sensitivity) square all plausible cases. We assess that the truth lies much closer to a cost which is independent of the oil price than fully proportional to it, hence we use a fixed cost as our baseline. This is since a majority of the total refining costs are capital costs, and for the remaining operating costs the fixed part (maintenance, labor) is usually somewhat larger than the variable operating costs (which include energy) (Sannan 2017). Similarly, transport costs very much depend of the cost of capital.

### 7.2.3 Price decoupling Urals and Brent

One noteworthy evolution is that the main oil price that Russia faces (Urals) has, since the start of the invasion, been somewhat disconnected from other oil prices (e.g., Brent). This could affect our results if a 1 USD price increase in Brent does not result in an equivalent increase in Urals. We investigate this empirically. Table 10 shows the results of four regressions, (1) and (2) are before the invasion and (3) and (4) after.[13] (1) and (3) regress the price level of Ural on Brent. There is a statistically insignificant drop in the estimate in the post-invasion period. The way the regression is specified, these estimates imply that a change in Brent implies a larger change in Urals. What is clearly significant is the difference in the constant which captures that Urals has been traded with a roughly 30 USD discount after the invasion.

*Table 10. Brent and Urals price and price first-difference.*

|  | Pre-invasion | | Post-invasion | |
| --- | --- | --- | --- | --- |
|  | (1) Brent price | (2) Brent FD | (3) Brent price | (4) Brent FD |
| Ural price | 0.947*** (0.00439) |  | 0.807*** (0.172) |  |
| Ural FD |  | 0.732*** (0.0192) |  | 0.881*** (0.156) |
| Constant | 4.269*** (0.278) | 0.00209 (0.0288) | 34.66** (16.28) | 0.854 (0.805) |
| Observations | 925 | 924 | 24 | 24 |
| Note: * p<0.1, ** p<0.05, *** p<0.001. Standard errors in parenthesis. First difference (FD) calculated as $price_t - price_{t-1}$. | | | | |

Columns (2) and (4) are more interesting from our perspective as they look at whether changes in the Brent and Urals prices are correlated. They are clearly so and, if anything, more so after the invasion (though no statistically significant difference in estimates). From this we draw the conclusion that there is no reason to believe that rising global oil prices will not result in a virtually equal rise in the Urals price. Given the short post-invasion period (which means we cannot filter the variation between short and long-run shocks) it is hard to know whether this will change later on.

---

[13] Data from Investing.com (2022). Data time coverage given by availability of Urals data. Price regressions: July 12, 2018 - February 23 2022 for pre-period, and February 24-March 29, 2022 for post period. First difference regressions: July 12, 2018 - February 24, 2022 for post period and February 25-March 29, 2022 for post period.

Sannan, S. (2017) *Understanding the Cost of Retrofitting CO2 capture in an Integrated Oil Refinery*.

Severen, C. and van Benthem, A.A. (2022) 'Formative Experiences and the Price of Gasoline', *American Economic Journal: Applied Economics*, 14(2), pp. 256–284. doi:10.1257/app.20200407.

Tax Foundation (2021) 'Gas Taxes in Europe', *Tax Foundation*, 12 August. Available at: https://taxfoundation.org/gas-taxes-in-europe/ (Accessed: 7 April 2022).

Time (2015) *A Former Russian Troll Explains How to Spread Fake News*, *Time*. Available at: https://time.com/5168202/russia-troll-internet-research-agency/ (Accessed: 6 April 2022).

Tylaz (2022) 'What are the salaries of Russian and Ukrainian soldiers, in fact', *Tylaz*, 26 February. Available at: https://www.tylaz.net/2022/02/26/what-are-the-salaries-of-russian-and-ukrainian-soldiers-in-fact-dan-negru-confessed-rejoice/ (Accessed: 6 April 2022).

U.S. Department of the Treasury (2022) *Ukraine-/Russia-related Sanctions*, *U.S. Department of the Treasury*. Available at: https://home.treasury.gov/policy-issues/financial-sanctions/sanctions-programs-and-country-information/ukraine-russia-related-sanctions (Accessed: 6 April 2022).

Wachtmeister, H. (2020) *World oil supply in the 21st century : A bottom-up perspective*. Doctoral dissertation. Uppsala University. Available at: http://urn.kb.se/resolve?urn=urn:nbn:se:uu:diva-419144 (Accessed: 5 January 2022).

World Bank (2022) *Military expenditure (current USD) - Russian Federation*. Available at: https://data.worldbank.org/indicator/MS.MIL.XPND.CD?locations=RU (Accessed: 6 April 2022).24